\documentclass[prb,onecolumn,amssymb,amsfonts,amsmath,showpacs,preprint,a4paper,superscriptaddress]{revtex4}
\usepackage{graphicx}

\begin{document}
\title{Ultrafast magnetization dynamics of Gd(0001): Bulk vs. surface}

\author{Muhammad Sultan}
\altaffiliation{on leave from: National Centre for Physics,
Islamabad, Pakistan}\affiliation{Freie Universit\"{a}t Berlin,
Fachbereich Physik, Arnimallee 14, 14195 Berlin,
Germany}\affiliation{Universit\"at Duisburg-Essen, Fakult\"at
f\"ur Physik, Lotharstr.~1, 47048 Duisburg, Germany}

\author{Alexey Melnikov}
\affiliation{Freie Universit\"{a}t Berlin, Fachbereich Physik,
Arnimallee 14, 14195 Berlin,
Germany}\affiliation{Fritz-Haber-Institut der Max-Planck
Gesellschaft, Faradayweg 4-6, 14195 Berlin, Germany}

\author{Uwe Bovensiepen}
\email[] {uwe.bovensiepen@uni-due.de} \affiliation{Freie
Universit\"{a}t Berlin, Fachbereich Physik, Arnimallee 14, 14195
Berlin, Germany}\affiliation{Universit\"at Duisburg-Essen,
Fakult\"at f\"ur Physik, Lotharstr.~1, 47048 Duisburg, Germany}

\date{\today}

\begin{abstract}
Ultrafast laser-induced demagnetization of Gd(0001) has been
investigated by magneto-induced optical second harmonic generation
and the magneto-optical Kerr effect which facilitate a comparison
of surface and bulk dynamics. We observe pronounced differences in
the transient changes of the surface and bulk sensitive
magneto-optical signals which we attribute to transfer of
optically excited, spin-polarized carriers between surface and
bulk states of the Gd(0001) film. A fluence dependent analysis of
the bulk magnetization dynamics results in a weak variation of the
demagnetization time constant, which starts at about 700~fs and
increases by 10\% within a fluence variation up to 1~mJ/cm$^2$. We
compare these results with fluence dependent changes in the
transient energy density calculated by the two temperature model.
The determined characteristic times of excess energy transfer from
the electron system to the lattice, which is mediated by e-ph
scattering, range from 0.2 - 0.6~ps. Such a more pronounced
fluence dependent change in the characteristic time compared to
the observed rather weakly varying demagnetization times suggests
a more advanced description of the optically excited state than by
the two-temperature model.
\end{abstract}


\maketitle

\section{Introduction}

The understanding of spin dynamics in ferromagnets is widely based
on radio frequency spectroscopy \cite{farle_RPP98} and inelastic
scattering techniques employing photons \cite{hilleb_94} or
neutrons \cite{mook_JMMM83}. These experimental methods provide
frequencies and decay rates of magnetic excitations by the line
position and line width, respectively. More recently time domain
techniques that employ short laser pulses have been established in
a considerable number of laboratories and provide a complementary
approach to analyze spin dynamics
\cite{hilleb_03,hillebrands_JPhysD08,kirilyuk_RMP2010}. In
pump-probe experiments the magnetization dynamics is driven by an
external stimulus like a laser or magnetic field pulse and is
probed by a second, time-delayed laser pulse. Using femtosecond
laser pulses, which are nowadays routinely provided by Ti:Sapphire
laser systems, the field of femtosecond magnetization dynamics has
developed enormously. One reason for the interest in this approach
might be that it is not only complementary to the established
tools for magnetization dynamics, but that the non-equilibrium
state and the respective dynamical magnetization processes have
become accessible for systematic experiments
\cite{kirilyuk_RMP2010}.

A number of novel phenomena like laser-induced demagnetization on
sub-picosecond time scales \cite{beaurepaire_PRL96,
stamm_NatMat07}, laser-driven formation of ferromagnetic order
\cite{ju_PRL04}, and magnetization reversal by individual laser
pulses \cite{stanciu_PRL07} demonstrates the potential for spin
manipulation in solid state materials on ultrafast time scales. At
the same time, these phenomena challenge our understanding of
ferromagnetism and ask for an appropriate description of the
observed dynamics.

Beside the investigation of multi-constituent materials with a
potential for novel phenomena, it is essential to develop insight
into the more general laser-driven dynamics to establish this
understanding. Finally, the meanwhile well established
sub-picosecond demagnetization of elemental metallic ferromagnets
is still controversial because the underlying elementary
interactions remain under discussion. It has been shown recently
that descriptions based on phonon-mediated spin-flip scattering
(Elliott-Yafet type) \cite{steiauf_PRB09,koopmans_NatMat10},
electron-spin-flip scattering (Stoner excitation)
\cite{atxitia_PRB10}, and ballistic transport of spin-polarized
charge carriers \cite{battiato_PRL10} can be used to describe
rather similar experimental data of femtosecond demagnetization in
Ni films. In the phonon-mediated process the magnetic moment is
considered to be transferred directly to the lattice through
phonons with appropriate symmetry obeying angular momentum
conservation. In the electronic spin-flip scattering process the
magnetic moment is not transferred within the elementary
scattering process itself, but is dissipated by secondary low
energy spin excitations, which in their sum lead to
demagnetization \cite{atxitia_PRB10}. In the transport concept
\cite{battiato_PRL10} the spin polarization is basically taken out
of the ferromagnet into a substrate or, alternatively, out of the
experimental observation window. Therefore, the current
understanding is not conclusive and a discrimination of these
above processes is important at this stage.

Our efforts have concentrated on rare earth ferromagnets and in
particular on Gadolinium which is considered to be a model system
for a Heisenberg ferromagnet with a magnetic moment localized at
the ion core. We have employed in earlier studies time-resolved
surface sensitive techniques which are optical magneto-induced
second harmonic generation (SHG) and photoelectron spectroscopy,
to analyze the excited state and its relaxation. The response to
the laser excitation is characterized by a coherent phonon-magnon
mode localized at the surface and incoherent dynamics of optically
excited electron-hole pairs, lattice vibrations, and magnetic
excitations. A comprehensive overview is given in
Refs.~\cite{melnikov_JPD08,melnikov10}.

Here we present experimental results obtained by the femtosecond
time-resolved magneto-optical Kerr effect (MOKE) on Gd(0001) films
to analyze the magnetization dynamics in the bulk part of the Gd
film, which we compare with the surface sensitive SHG signal. We
find pronounced differences during the first 2~ps which we
attribute to transfer of spin-polarized carriers between the
surface and bulk part of the film. Moreover, we study the fluence
dependence of the demagnetization process and compare it with
predictions based on the two-temperature model.

\section{experimental aspects}

The experiments were carried out at a setup that combines an
ultrahigh vacuum (UHV) chamber and a femtosecond (fs) Ti:Sapphire
laser. For details see~\cite{bovensiepen_JPCM07}. Gd was
evaporated from an electron beam heated crucible under UHV
conditions onto a W(110) substrate at 300~K. Annealing to 700~K
subsequent to the deposition results in smooth epitaxial films
\cite{aspelmeier_jmmm94}. Here we studied films with 20~nm
thickness. Pump-probe experiments were performed at 50~K
equilibrium temperature of the sample which was cooled by a liquid
He cryostat. Laser pulses of 35~fs duration and 40~nJ energy were
generated by a cavity dumped Ti:Sapphire oscillator at 800~nm
central wave length with a repetition rate of 1.52~MHz. For
pump-probe experiments the laser output was divided at a ratio
4:1. To detect the p-polarized second harmonic (SH) generated by
the p-polarized probe pulse a monochromator selecting 400~nm and a
photomultiplier were used. To analyze the MOKE a balance detection
scheme was used to measure the polarization rotation of the probe
pulse $\theta$ in the longitudinal MOKE geometry. The
magneto-optical signals were analyzed by a lock-in amplifier for
MOKE and a single photon counter for SHG to detect the
differential signals for open and blocked pump in opposite
saturation magnetic fields
($H_{\mathrm{s}}^{\uparrow,\downarrow}$) as a function of
pump-probe delay $t$. Formally negative delays $t<0$ denote the
state before the optical excitation has occurred. For MOKE we show
below
\begin{equation}
\frac{ \Delta \theta(t)}{\theta_0} = \frac{ \theta(t)}{\theta_0}
-1 = \frac{\vartheta(H_{\mathrm{s}}^{\uparrow},t)-
\vartheta(H_{\mathrm{s}}^{\downarrow},t)}{\vartheta(H_{\mathrm{s}}^{\uparrow},t<0)-
\vartheta(H_{\mathrm{s}}^{\downarrow},t<0)}-1
\end{equation}
where $\vartheta$ is the angle between the linear polarization of
the reflected probe pulse and a fixed reference polarization,
$\theta$ is the magneto-optical Kerr rotation. $\Delta \theta$ is
the time-dependent change in the Kerr rotation which represents
the magnetization of the bulk part of the Gd film.

SHG is detected in the transversal geometry leading to
magneto-induced changes in the SH intensity.
\begin{equation}
I_{2\omega}^{\uparrow,\downarrow} \propto
\left(E_{2\omega}^{\mathrm{even}}\right)^2+\left(E_{2\omega}^{\mathrm{odd}}\right)^2\pm
2E_{2\omega}^{\mathrm{even}} E_{2\omega}^{\mathrm{odd}}\cos{\phi}.
\end{equation}

Here $E_{2\omega}^{\mathrm{even}}$ and
$E_{2\omega}^{\mathrm{odd}}$ are the SHG optical fields which
behave as even or odd with respect to reversal of the
magnetization $M$. The phase between these two field contributions
is $\phi$ which is smaller than $15^\circ$ and weakly
time-dependent \cite{melnikov_JPD08}. Since
$E_{2\omega}^{\mathrm{odd}}\propto M$ we plot below

\begin{equation}
\Delta_{2\omega}^{\mathrm{odd}}(t)\approx
\frac{E_{2\omega}^{\mathrm{odd}}(t)-E_{2\omega}^{\mathrm{odd}}(t<0)}{E_{2\omega}^{\mathrm{odd}}(t<0)}
\approx \frac{M(t)-M(t<0)}{M(t<0)}=\frac{\Delta M(t)}{M_0}
\end{equation}

which represents the time-dependence of the surface sensitive
magneto-optical signal.

\section{Experimental Results and Discussion}

The article presents results of time-resolved SHG and MOKE. The
first section focuses on the comparison of surface and bulk
sensitive detection of the transient magnetic state in Gd(0001).
The second section contains a detailed study of time-resolved MOKE
and reports on the fluence dependence of the bulk demagnetization.

\subsection{Surface and bulk demagnetization dynamics}

\begin{figure} \centering
 \includegraphics[width=0.49\columnwidth]{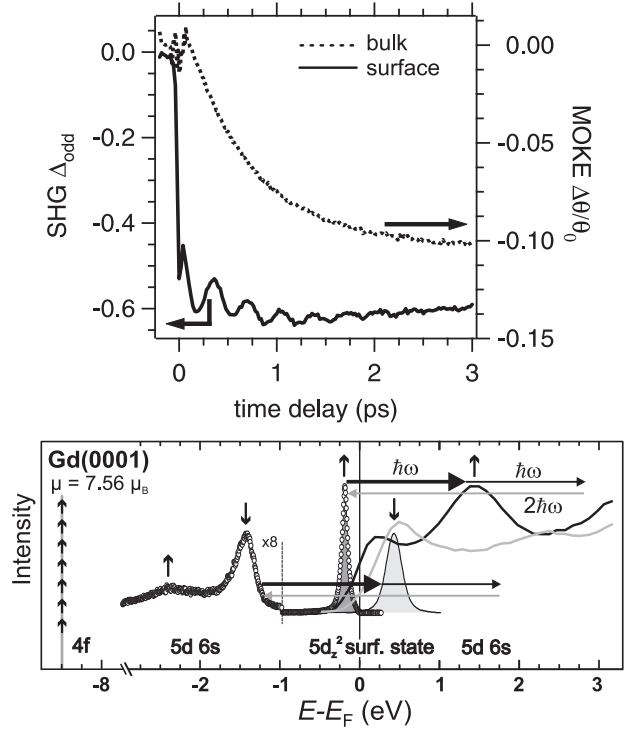}
\caption{\label{Fig1}Top panel: Time-dependent magneto-optical
signals measured on 20~nm thick Gd(0001) films, which were grown
epitaxially on W(110). The solid line shows the pump-induced
change of the magnetic SHG contribution
$\Delta^{\mathrm{odd}}_{2\omega}$ sensitive to the surface.
Similar results were published before
\cite{melnikov_PRL03,loukakos_PRL07,melnikov_JPD08}. The dotted
line depicts the pump-induced MOKE polarization rotation $\Delta
\theta$ normalized to the static rotation $\theta_0$. The bottom
panel is reprinted from Ref.~\cite{melnikov_JPD08} and shows the
valence electron states of Gd(0001) taken from normal direction
photoemission (circles), inverse photoemission (solid lines), and
scanning tunneling spectroscopy data above the Fermi level (solid
line, filled) \cite{rehbein_PRB03}. The exchange-split surface
state (filled area) appears around the Fermi level. Vertical
arrows represent the majority and minority character of the
electronic states. Indicated are the two main absorption channels
for 1.5~eV pump photons and the resonant second harmonic probing
scheme.}
\end{figure}

We start by comparing the transient SHG and MOKE signals that
probe the surface and bulk of the Gd(0001) film, respectively. In
Fig.~\ref{Fig1}, top panel, we show
$\Delta^{2\omega}_{\mathrm{odd}}$ (solid line, left axis, surface)
and $\Delta \theta / \theta_0$ (dotted line, right axis, bulk)
taken at nominally identical experimental conditions. Both signals
indicate a pronounced laser-induced demagnetization, but their
transient behavior is very different. The magnetic SHG signal is
reduced within the laser pulse duration of 35~fs by 0.5 and
exhibits on that level oscillations that are damped out within
3~ps. The signal starts to recover after 1.5~ps and returns to its
value before optical excitation after several 100~ps (not shown,
see Ref.~\cite{liso_PRL05}). The oscillatory part has been
explained before as a coupled phonon-magnon mode localized to the
surface \cite{melnikov_JPD08,melnikov10}. It is therefore
expected, that it is not found in the bulk signal. The MOKE data
can be described by a continuous reduction that will be fitted by
a single exponential decay below and tends to saturate at a delay
of 3~ps, before it demagnetizes further, not shown see Ref.~
\cite{wietstruk_subm10}. Overall, the bulk sensitive signal is
reduced on a much slower time scale than the surface one and is in
agreement with femtosecond x-ray magnetic dichroism studies
\cite{wietstruk_subm10} if the different excitation densities are
considered. We therefore conclude that the time-resolved MOKE
signal probes the time-dependent magnetization of the film. We
note here that a detailed comparison of MOKE ellipticity and
rotation points to time-dependent variations of the
magneto-optical constants at delays where the excess energy
resides predominantly in the electronic system, as will be
discussed in a forthcoming publication \cite{sultan_10}.

To explain the difference between the surface and bulk
demagnetization it is informative to inspect the electronic
structure at the surface and to take transfer processes into
account. Fig.~\ref{Fig1}, bottom, depicts several data sets which
in their combination represent the electronic valence states of
epitaxial Gd(0001) films. For a ferromagnetically ordered
situation the states are exchange-split due to intra-atomic
exchange interaction with the strong magnetic moment of the half
filled $4f$ shell. At the $\overline{\Gamma}$-point the bulk
$5d$-states appear at 1.4~eV (minority) and 2.4~eV (majority)
binding energy and disperse towards $E_{\mathrm{F}}$ with
increasing $k_{||}$ \cite{kurz_JPCM02}. The majority component of
the exchange-split $5d_{z^2}$-surface state is dominantly occupied
and the minority one unoccupied. The system also contains
unoccupied exchange-split bulk states \cite{donath_PRL96}.

For linear polarized light, within the dipole approximation,
optical transitions proceed among occupied and unoccupied
electronic states in (bulk) valence bands with the same spin
character. In case of the excitation by the 800~nm (or 1.55~eV)
pump laser pulses, an additional excitation channel becomes
available at the surface because of resonant transitions coupling
the bulk and the localized surface electronic states, see
Fig.~\ref{Fig1}, bottom. Therefore a difference of the surface
sensitive SHG response with respect to the bulk sensitive MOKE can
be expected. Similarly to a charge-transfer excitation across
molecular interfaces \cite{yang_PRB09}, such surface-bulk
resonances lead to effective electron transfer between surface and
bulk which modify the transient population of the minority and
majority components of the surface state. The population of the
initially occupied majority surface state component is reduced due
to the surface-to-bulk spin-up electron transfer, see
Fig.~\ref{Fig1}, bottom. At the same time, the population of the
initially unoccupied minority surface state component increases
due to the bulk-to-surface spin-down electron transfer, which can
alternatively be viewed as a surface-to-bulk spin-down hole
transfer. The carriers in the surface state which are transfered
to bulk bands represent initially a wave packet at the surface
which spreads and propagates into the bulk material with its Fermi
velocity $v_{\mathrm{F}}$ of about 1 nm/fs~\cite{brorson_RPL87}.
Thus, a spin-polarized current which propagates from the surface
to the bulk is optically excited. In the vicinity of the surface
this current propagates ballistically and consists of electron and
hole contributions. Spin-down holes have the same spin
polarization as spin-up electrons. Therefore, the spin components
of these two contributions will add up. If the charge of electrons
and holes is in sum zero we conclude that a net spin current
between surface and bulk is excited. It is likely that the
efficiency of electronic transitions in spin-up and spin-down
channels are different (Fig.~\ref{Fig1}, bottom) which will lead
to a prevalence of either the electron or hole component of the
spin-polarized current. Note that this does not necessarily imply
charging of the surface because a transient charge imbalance will
be screened by electron rearrangement on the time scale of the
inverse plasma frequency \cite{borisov_CPL04}. Note that on the
basis of these considerations an effect due to optically excited
transfer of spin polarization between surface and bulk should
proceed within few femtoseconds, i.e. well within the time
resolution of the SHG experiment.

As seen in Fig.~\ref{Fig1}, top panel, the pronounced initial drop
in the surface sensitive magneto-optical signal occurs within the
experimental time resolution and clearly faster than the bulk
demagnetization. Therefore, we explain the pronounced difference
in the bulk and surface demagnetization times by spin transfer
between surface and bulk of the film. The time scale at which the
magnetic SHG signal $\Delta_{\mathrm{odd}}^{2 \omega}$ changes
from 0 to -0.5 at time zero is in agreement with a ballistic
character of these transport effects.

The pump-induced reduction of the relative magneto-optical signals
differ by a factor of six. We consider the resonant surface
excitation among bulk and the surface states near
$\overline{\Gamma}$, see Fig.~\ref{Fig1}, as essential for an
explanation of the pronounced effect at the surface.

\subsection{Fluence dependent demagnetization dynamics in the bulk}

\begin{figure} \centering
\includegraphics[width=0.49\columnwidth]{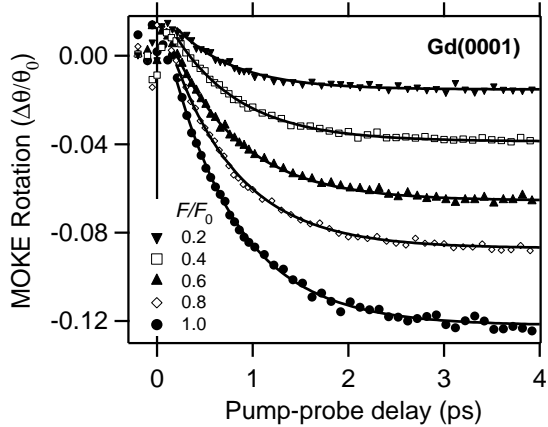}
\caption{\label{Fig2}Symbols indicate experimental data of
time-resolved MOKE for different relative pump fluences $F/F_0$,
with $F_0 \approx 1$~mJ/cm$^2$. The solid lines are fits
considering a single exponential decay.}
\end{figure}

The setup used for the present study allowed in addition a more
detailed analysis of the bulk demagnetization phenomenon. Using a
combination of half wave plate and Glan-Thomson polarizer the
absorbed fluence has been reduced step wise from the maximum value
$F/F_0=1$, $F_0$ was determined to be $1.0\pm0.3$~mJ/cm$^2$.
Fig.~\ref{Fig2} shows representative time-resolved MOKE curves for
different $F/F_0$ up to delay times of 4~ps. The values at 4~ps
decrease linearly with $F/F_0$. This ensures that we are analyzing
a low excitation density regime reasonably far away from a full
demagnetization of the sample, where magnetic fluctuations would
contribute to the ultrafast magnetization dynamics
\cite{atxitia_PRB10,sultan_10}. Near time zero the data feature a
less clear behavior and an effectively positive contribution in
$\Delta \theta/\theta_0$. The time delay at which the signal
crosses the zero line shifts with increasing fluence closer to
time zero. While for $F/F_0=0.2$ the amplitude of the positive and
negative contributions are comparable to each other, the negative
one dominates for larger $F/F_0$. In this study we focus on the
pronounced demagnetization dynamics, i.e. at the fluences where
the negative contribution dominates. The origin of the positive
contributions are currently under discussion since they could
originate from transient changes in the magneto-optical constants
or the spin polarization of the $5d$ electrons.

\begin{figure}
\includegraphics[width=.49\columnwidth]{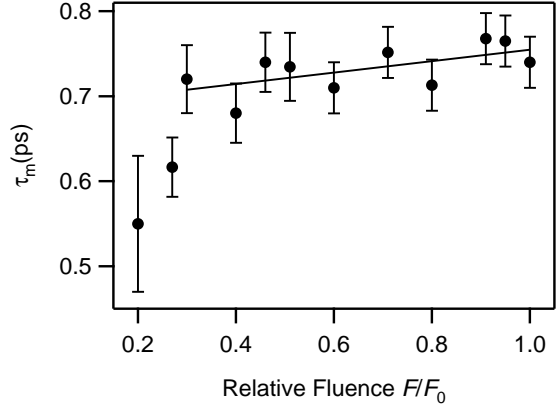}
\caption{\label{Fig3}Characteristic times determined from the
decay times in the single exponential fitting of $\Delta \theta /
\theta_0$ as a function of relative fluence. The line is a linear
fit to the data with $F/F_0\geq0.30$}
\end{figure}

All data were fitted by a single exponential time dependence with
fixed time zero and variable $\Delta \theta / \theta_0$ at 0 and
4~ps. The obtained fits are shown in Fig.~\ref{Fig2} and describe
the experimental data well. Fig.~\ref{Fig3} depicts the time
constants $\tau_{\mathrm{m}}$ determined by the fitting procedure.
The numbers range from about 0.5 to 0.8~ps with a trend towards
larger times for higher $F/F_0$. The smallest $\tau_{\mathrm{m}}$
are obtained for the lowest $F/F_0=0.20$ and 0.27 and the
respective $\tau_{\mathrm{m}}$ are out of the weak linear increase
observed for $F/F_0\geq 0.30$. We already mentioned that for such
small $F/F_0$ the positive signal near time zero is comparable in
size to the demagnetization observed at later delays. It is
therefore well possible that the obtained $\tau_{\mathrm{m}}$ is
influenced by the processes that are responsible for the positive
$\Delta \theta / \theta_0$ and we refrain from further conclusions
based on $\tau_{\mathrm{m}}$ obtained for $F/F_0 < 0.30$.

We fitted the values for $F/F_0\geq 0.30$ by a linear dependence.
The result is shown in Fig.~\ref{Fig3} by a line. We find that the
demagnetization time increases by $67\pm30$~fs within $F/F_0=1$
with a nominal zero fluence limit of
$\tau_{\mathrm{m}}=690\pm20$~fs.

For itinerant ferromagnets like Ni or Co an observed increase in
$\tau_{\mathrm{m}}$ with fluence in combination with a finite
value at zero fluence was explained by model descriptions based on
Elliott-Yafet scattering \cite{koopmans_NatMat10} and Stoner
excitations \cite{atxitia_PRB10}. The case of Gd has been
discussed in Ref.~\cite{koopmans_NatMat10}. It features two
separate demagnetization times in agreement with the experimental
observation \cite{wietstruk_subm10}. These two time scales are
attributed (i) to cooling of the hot electron system through e-ph
coupling for the fast time scale, which is discussed in the
present article, and (ii) to the demagnetization determined by the
equilibrium spin-flip probability, which is weaker in Gd than for
$3d$ transition metal ferromagnets and sets the slower timescale
observed in Refs.~
\cite{vaterlaus_PRL91,melnikov_PRL08b,wietstruk_subm10}.

\begin{figure}
\includegraphics[width=.49\columnwidth]{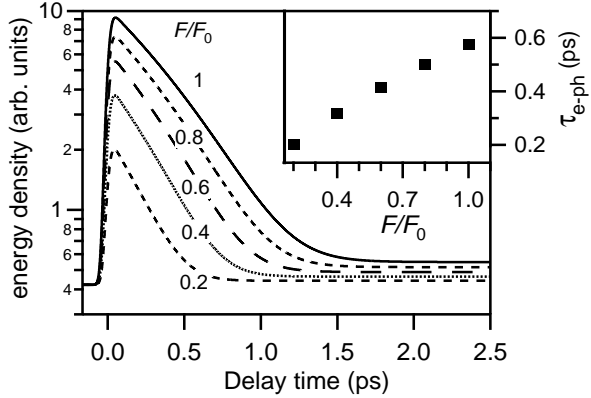}
\caption{\label{Fig4}Main panel: Time-dependent excess energy
density for different relative fluence, which was calculated by
the two-temperature model. The inset shows the characteristic
times $\tau_{\mathrm{e-ph}}$ of energy transfer from the
electronic system to the lattice mediated by electron-phonon
scattering which was determined by single exponential decay to the
excess energy density in time interval from 100 to 800~fs.}
\end{figure}

This description predicts a fluence dependence of the ultrafast
demagnetization of Gd that is determined by the increase of the
transient lattice temperature $T_{\mathrm{l}}$. We have calculated
$T_{\mathrm{e}}(t)$ and $T_{\mathrm{l}}(t)$ for different fluence
by the well known two-temperature model \cite{anisimov_sovphys74}
in refined versions, which were published earlier
\cite{groeneveld_PRB95,bonn_PRB00}; regarding its application to
Gd(0001) see \cite{bovensiepen_JPCM07}. Considering a variation in
fluence following the experimentally investigated range we find a
pronounced variation of the time scale on which
$T_{\mathrm{e}}(t)$ and $T_{\mathrm{l}}(t)$ equilibrate (not
shown). This change in time scale is at a first glance more
pronounced than the changes found for the demagnetization time in
Fig.~\ref{Fig3}. However, a systematic analysis of the time scale
at which $T_{\mathrm{e}}(t)$ and $T_{\mathrm{l}}(t)$ are changed
is non-trivial since they vary in a non-exponential way.
Therefore, we turn to a discussion of the excess energy $\epsilon$
of the electronic system which is related to the electron
temperature through $\epsilon= \gamma T_{\mathrm{e}}(t)^2$, with
$\gamma$ being the linear parameter in the temperature dependent
specific heat of the electron system. Fig.~\ref{Fig4} shows the
results obtained for $\epsilon(t)$ with $F/F_{{\mathrm{0}}}$
varying between 0.2 and 1. The transient behavior can be described
by a single exponential that represents energy transfer from the
electron system to the lattice. The exponential time scales were
determined with 100--800~fs and are plotted in the inset of
Fig.~\ref{Fig4}. The energy transfer time shifts to larger values
for higher fluence and changes by three times in the investigated
fluence range. Consequently the energy content of the lattice
changes with these varying time constants.

The comparison between time-dependent demagnetization curves and
the time evolution of excess energy emphasizes two points which
lead to the conclusion that the explanation of the ultrafast
demagnetization in Gd solely through the time-dependent electron
or lattice temperature is incomplete. At first we argue that the
range of the energy transfer times $\tau_{\mathrm{e-ph}}$ found in
the investigated fluence range is with 0.2--0.6~ps considerably
broader than the range observed for $\tau_{\mathrm{m}}$ which is
about 0.65--0.8~ps (see Figs.~\ref{Fig3},\ref{Fig4}). Second, and
maybe more general, the demagnetization can be described by a
simple exponential time dependence similar to the transient energy
density. In contrast, the transient electron and lattice
temperatures follow a more complicated evolution. Moreover, spin
fluctuations which present a sizeable energy content of the system
in particular in excited magnetic systems \cite{dankov_PRB98},
might be essential to take into account.

\subsection{Discussion}

Our pump-probe analysis of fs laser-induced magnetization dynamics
in Gd(0001) revealed two qualitatively different mechanisms, which
both change transiently the magnetic moment per atom. The one
mechanism is transfer of angular momentum from the spin system to
another degree of freedom of the sample, which is finally the
lattice. In this case the magnetic moment of the sample as a whole
is reduced. Such mechanisms are discussed in the literature
\cite{koopmans_NatMat10,atxitia_PRB10} to offer consistent
descriptions of ultrafast demagnetization as observed in several
time-resolved magneto-optical studies, see e.g. Ref.~
\cite{beaurepaire_PRL96}, and x-ray magnetic-circular dichroism
(XMCD) experiments \cite{stamm_NatMat07} for Ni.  The observed
demagnetization time of about 700~fs determined for Gd(0001) can
be explained by relaxation of the optically excited electrons by
interactions with phonons or spin waves which subsequently
interact with the lattice. On a more microscopic level the
demagnetization process in Ni and Gd must be different, because in
a transition metal the magnetic moment is formed in the same band
which is also optically excited. In a lanthanide like Gd the
dominant part of the magnetic moment is localized near the ion
core and resides in the $4f$ level, which is not primarily
optically excited \cite{wietstruk_subm10}. To achieve significant
demagnetization the excess energy generated by the laser
excitation must be transfered to the $4f$ electrons. While details
of this process will be discussed in a forthcoming publication, we
note here that optically excited $5d$ electrons are coupled to the
localized $4f$ electrons through intraatomic exchange interaction.
Ultrafast changes of the magnetization, which are investigated by
MOKE, can be expected to be dominated by the $5d$ contribution to
the magnetic moment. The femtosecond XMCD study performed at the
Gd M$_5$ edge probes the magnetic moment of the $4f$ levels
directly \cite{wietstruk_subm10}. A comparison of the time scales
of demagnetization obtained in both these experiments suggests
that the $4f$ and $5d$ contribution to the magnetic moment are
essentially strongly coupled on ultrafast time scales accessible
here and demagnetize as a total magnetic moment.

The second mechanism investigated here is spin transfer which we
probe by means of the exchange-split spin polarized 5$d_{z^2}$
surface state of Gd(0001). Here the magnetic moment integrated
over the sample remains constant, but is redistributed between
different parts of the sample, in our case among surface and bulk
electronic states. We explain this process to originate from the
optical transitions that couple the exchange-split surface and
bulk states. In the literature a similar scenario has been
discussed recently as a potential explanation for bulk
demagnetization in Ni \cite{battiato_PRL10}. In this theoretical
study the authors consider the optically excited transport of hot
carriers in metallic samples which redistribute the magnetic
moment from the ferromagnetic layer into a para- or diamagnetic
substrate. Such transport effects of hot carriers are known since
early investigations of femtosecond electron dynamics in metallic
layers \cite{brorson_RPL87}, however, to what fraction they are
indeed responsible for the ultrafast demagnetization remains an
interesting question which is to be clarified in future
investigations.

\section{Conclusions}

By combining ultrafast bulk and surface sensitive magneto-optical
techniques we have analyzed the femtosecond laser-induced
demagnetization of epitaxial Gd(0001) films. In the bulk the
demagnetization occurs with a characteristic time of about 0.7~ps.
Considering that the surface sensitive signal changes within the
laser pulse duration of 35~fs and taking resonant optical
transitions between valence electronic surface and bulk states
into account, we attribute this ultrafast change to transfer of
spin-polarized charge carriers between surface and bulk states.
Variation of the pump fluence up to 1~mJ/cm$^2$ shows a weak
increase in the bulk demagnetization time of about
70~fs/(mJ/cm$^2$). A comparison with the fluence dependence
expected from the energy transfer among electrons and the lattice
revealed that albeit there is qualitative agreement in the
increasing trend of demagnetization time such description remains
incomplete due to pronounced quantitative variations.

\begin{acknowledgments}We are grateful for the continuous support by M.
Wolf and acknowledge fruitful discussions with O.
Chubykalo-Fesenko. This work was supported by the Deutsche
Forschungsgemeinschaft through ME3570/1 and by the HEC-DAAD.
\end{acknowledgments}


\bibliographystyle{prsty}

\end{document}